# A Novel RIS-Aided EMF Exposure Aware Approach using an Angularly Equalized Virtual Propagation Channel

Nour Awarkeh*, Dinh-Thuy Phan-Huy† and Marco Di Renzo*‡

*CentraleSupélec, France, †Orange Innovation, Châtillon, France, ‡CNRS, Université Paris-Saclay, France

*Abstract*—Massive Multiple-Input Multiple-Output systems with beamforming are key components of the 5$^{th}$ and the future 6$^{th}$ generation of networks. However, in some cases, where the BS serves the same user for a long period, and in some propagation conditions, such systems reduce their transmit power to avoid creating unwanted regions of electromagnetic field exposure exceeding the regulatory threshold, beyond the circle around the BS that limits the distance between people and the BS antenna. Such power reduction strongly degrades the received power at the target user. Recently, exposition aware beamforming schemes aided by self-tuning reconfigurable intelligent surfaces derived from maximum ratio transmission beamforming, have been proposed: truncated beamforming. However, such scheme is highly complex. In this paper, we propose a novel and low complexity reconfigurable intelligent surface aided beamforming scheme called Equalized beamforming, which applies maximum ratio transmission to an angularly equalized virtual propagation channel. Our simulations show that our proposed scheme outperforms the reduced beamforming scheme, whilst complying with the exposition regulation.

*Keywords*— Massive MIMO, electromagnetic field exposure, maximum ratio transmission, beamforming, reconfigurable intelligent surfaces.

## I. INTRODUCTION

Massive Multiple-Input Multiple-Output (MMIMO) systems and adaptive beamforming (BF) key components of the 5$^{th}$ generation and the future 6$^{th}$ generation (6G) of networks, as they enable to deliver high throughput [1][2]. As an example, a Base Station (BS) equipped with a MMIMO antenna and transmitting data at full power $\chi^{max}$ maximizes the received power and the delivered data rate at the target user equipment (UE) by using *Maximum Ratio Transmission (MRT) BF* scheme [3][4]. However, the regulation fixes a maximum electromagnetic field (EMF) exposure (EMFE) threshold that cannot be exceeded outside a predefined region, for instance, a limit circle in the case where there is no obstacle around the BS), on a pre-defined time-average [5]. Thanks to the use of MRT BF and MMIMO, the shape of the area where the threshold is exceeded changes with the target user being scheduled. Hence, the threshold may be exceeded only temporarily, without violating the regulation [5]-[10]. However, in the case where the BS must serve the same target UE for a long period (which is true for fixed wireless access backhaul for instance), it is necessary to contain the over-exposed area (i.e., the area where the threshold is exceeded) inside the circle, all the time, even in its strongest directions. As illustrated in Fig.1-a), without any power control, the over-exposed area due to MRT BF scheme exceeds the limit circle in directions corresponding to strong propagation paths between the antenna and the current target UE. Such scheme cannot be deployed. Also, we foresee that even more stringent thresholds with arbitrarily larger limit circles could be requested in the future by some cities. One first and simple solution [11] consists in keeping using MRT BF and transmitting with a reduced power $\chi^{red} < \chi^{max}$ at the BS, so that the entire over-exposed area remains inside the circle, all the time, even in its strongest directions. Unfortunately, as illustrated in Fig. 1-b), the received power and throughput at the target UE is degraded by such *Reduced BF* scheme.

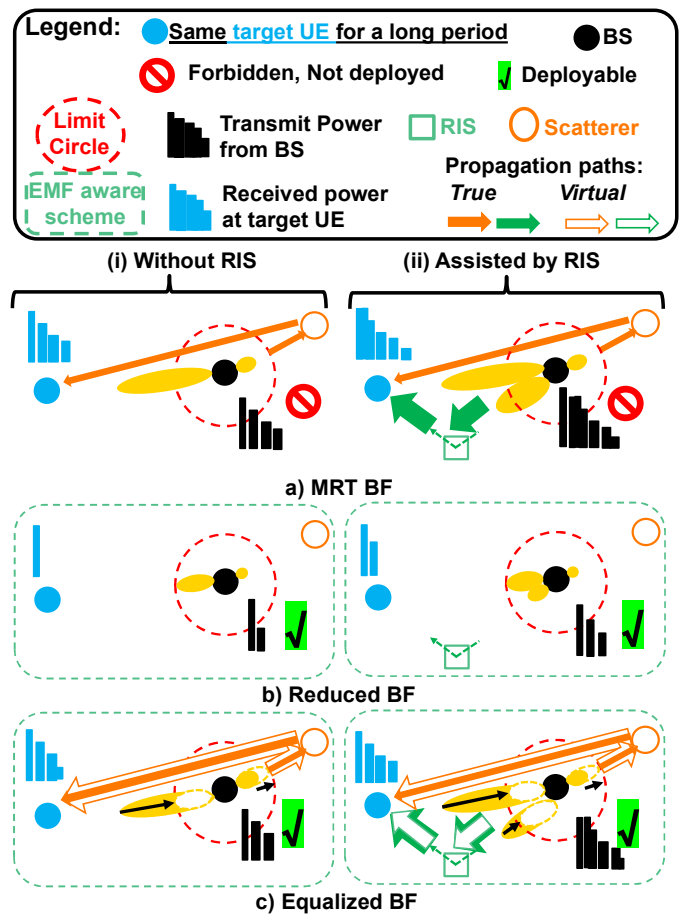

Fig. 1. Visualisation of studied BF schemes and corresponding over-exposed areas, assuming the same target UE is served for a long period.

A second solution [11], to overcome the drawback of Reduced BF, named *Truncated BF* has been proposed. It truncates the MRT BF radiation pattern, only in some directions: the directions where the over-exposed area exceeds the limit circle. The Truncated BF scheme does not impact directions already inside the circle. Compared to the Reduced BF scheme, the Truncated BF scheme uses a transmit power $\chi^{trunc} > \chi^{red}$, and thus delivers data to the target UE with an improved received power, whilst remaining compliant with the EMFE constrain. [11] has also exploited the recent concept of smart radio environments for 6G, where the propagation environment itself (according to some Channel State Information (CSI)) is reconfigured thanks to the use of reconfigurable intelligent surfaces (RISs) [12]-[16], to further improve the performance of the aforementioned BF schemes. Note that [11] uses RISs for EMFE reduction due to the BS in the downlink, whereas [15][16] use RISs to reduce self-EMFE reduction due to the smartphone in the uplink. More precisely, in [11], a RIS with sensing capability, and continuous (instead of discrete) phase-shifting capability (such as the prototypes of [17][18]) is considered. The RIS first measures the propagation channel between itself and the target UE. Then, the RIS reconfigures itself to 'turn itself electronically' into the direction of the target UE. Several RISs are deployed in the environment and self-tuned. Finally, all the BF schemes derived from MRT, by applying MRT to such pre-shaped propagation environment, all naturally spread their radiation patterns in additional directions (the directions of the RISs), as illustrated in Fig. 1-a) and 1-b).

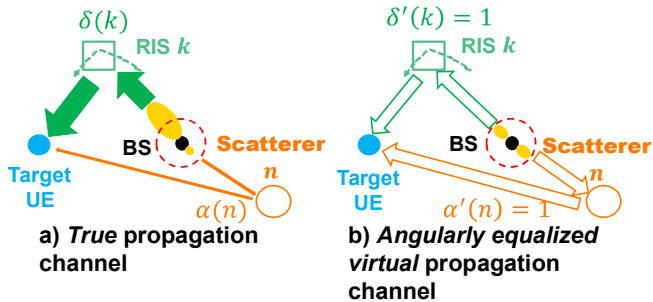

Fig. 2. True (a) and angularly equalized virtual (b) propagation channels

However, whereas the Reduced scheme is rather simple, the Truncated scheme is complex [11], as it requires 1) the projection of the MRT BF onto a Discrete Fourier Transform codebook, 2) The identification of all directions exceeding the limit circle, 3) The truncation of each of these directions. Another drawback of the Truncated BF scheme is that it leaves directions inside the limit circle unchanged, whereas they could be exploited to transmit more energy towards the target UE.

In this current paper, we propose a new BF scheme called *Equalized BF*, illustrated in Fig. 1-c), where the BS computes its BF precoder as follows:

- The BS estimates the directions of departure of the *true* propagation channel (for instance thanks to channel sounding techniques in [19]-[21]) illustrated in Fig. 2-a).
- The BS computes an *angularly equalized virtual* propagation channel, which paths have the same directions of departure as in the true channel, but with all the gains being equalized to 1, as illustrated in Fig. 2-b).
- The BS computes the precoder by applying MRT to the angularly equalized virtual propagation channel instead of the true channel.
- The BS computes the transmit power to guarantee that the EMFE constrain is met on the limit circle.

As for [11], we exploit self-tuned RISs that are electronically turned in the direction of the target UE, and the BS is not even aware that there is one or several RIS(s). Such novel approach based on angularly equalized virtual propagation channel is expected to ensure that in all directions of propagation, the over-exposed area exactly meet the limit circle, in a simple manner. The paper is organized as follows. In Section II, we describe the system model. In Section III, we visualize the impact of the BF schemes on the over-exposed area for a given propagation channel sample. In Section VI, we quantify and compare the performance of each scheme based on statistics over randomly generated propagation channels. Finally, we give some concluding remarks in Section V. The following notations are used throughout the paper: $j^2 = -1$, if $x \in \mathbb{C}$, then $\arg(x)$ and $|x|$, are the argument and the module, of $x$, respectively. If $\mathbf{A} \in \mathbb{C}^{M \times N}$, $\mathbf{A}_{m,n}$ is the coefficient of line $m$ and column $n$, with $1 \leq m \leq M$ and $1 \leq n \leq N$, $\|\mathbf{A}\| = \sqrt{\sum_{m=1}^{M} \sum_{n=1}^{N} |\mathbf{A}_{m,n}|^2}$, $\mathbf{A}^\dagger$ and $\mathbf{A}^T$ are the Hermitian and transpose of $\mathbf{A}$. If $\vec{a}, \vec{b} \in \mathbb{R}^{3 \times 1}$ are two points vectors in space with cartesian coordinates, then $\vec{a} \cdot \vec{b}$ is their scalar product. $E[.]$ is the expectation operation.

## II. SYSTEM MODEL

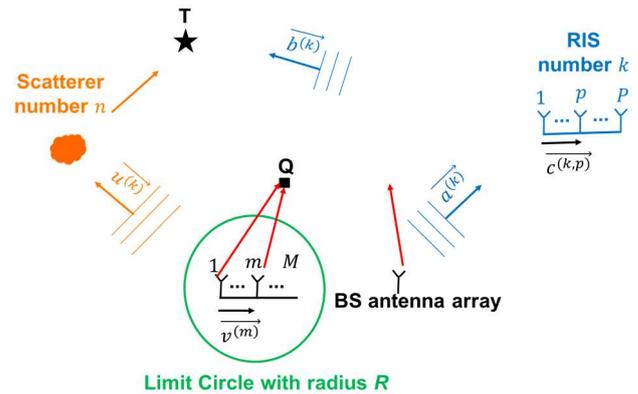

Fig. 3. Propagation channel model

We consider the downlink communication between a BS and a target UE. The BS is equipped with an MMIMO antenna consisting of a uniform linear array of $M$ antenna elements spaced by $0.5\lambda$, where $\lambda$ is the wavelength. The target UE has a single antenna. We assume that $K \geq 0$ RIS(s) are deployed in the environment. Note that $K = 0$, corresponds to a scenario without RIS assistance. We assume that each RIS is a uniform linear array of $P$ elements spaced by $0.5\lambda$. Finally, $T \in \mathbb{R}^{3\times 1}$ denotes the target UE location, $Q$ denotes a position close to the BS, $\omega^{threshold}$ denotes the threshold of received power that must not be exceeded to remain compliant with the EMFE constrain, $\mathcal{C}$ denotes the limit circle around the BS for EMFE, $R$ denotes the radius of $\mathcal{C}$. % denotes percent.

*A. Propagation Channel Model*

As illustrated in Fig. 3, we consider a multipath propagation environment between the BS and the target UE, with $N > 1$ scatterers. We assume that an Orthogonal Frequency Division Multiplexing (OFDM) waveform is used and restrict our analysis to a single sub-carrier of the OFDM waveform, as our analysis can be easily generalized to any other sub-carrier of the multi-carrier waveform. With this latter assumption, the propagation channels can be modeled with complex matrices or vectors and the following channel matrices are defined: $\mathbf{s}, \mathbf{h}, \mathbf{g} \in \mathbb{C}^{1 \times M}$ model the multipath propagation channels between the BS and the target UE, *through the scatterers*, *through the RIS(s)*, *through the scatterers and RIS(s)) together*, respectively. $\mathbf{q}(Q) \in \mathbb{C}^{1 \times M}$ models the propagation channel between the BS and a UE Q close to (and in line-of-sight of) the BS. With these notations, $\mathbf{g}$ is given by:

$$\mathbf{g} = \mathbf{s} + \mathbf{h}. \quad (1)$$

Note that in the absence of RIS, $\mathbf{h}$ is the null vector and $\mathbf{g} = \mathbf{s}$. The expressions of $\mathbf{s}$, $\mathbf{h}$ and $\mathbf{q}(Q)$ are provided hereafter.

The scatterers are assumed to be located far from the BS and from the target UE. Hence, the planar wave approximation applies. With this assumption, $\mathbf{s} \in \mathbb{C}^{1 \times M}$ is given by:

$$\mathbf{s}_m = \sum_{n=1}^{N} \alpha^{(n)} e^{j\frac{2\pi}{\lambda}\left(\overrightarrow{u^{(n)}} \cdot \overrightarrow{v^{(m)}}\right)}, 1 \leq m \leq M, \quad (2)$$

where, $\alpha_n$ is the gain (modeled as a complex random gaussian variable under the Rayleigh fading assumption, with $\mathrm{E}\left[\left|\alpha^{(n)}\right|^2\right] = 1$) of the path passing by the $n^{th}$ scatterer, $\overrightarrow{u^{(n)}}$ is the unitary vector indicating the BS-to-$n^{th}$ scatterer direction, and $\overrightarrow{v^{(m)}}$ is the vector between the positions of the 1st and the $m^{th}$ element of the BS.

When RIS are deployed in the environment (i.e., when $K > 0$), RISs as well, are assumed to be far from the BS and from the target UE. Hence, again, the planar wave approximation applies. With these assumptions, $\mathbf{h} \in \mathbb{C}^{1 \times M}$ is given by:

$$\mathbf{h}_m = \sum_{k=1}^{K} \delta^{(k)} e^{j\frac{2\pi}{\lambda}\left(\overrightarrow{a^{(k)}} \cdot \overrightarrow{v^{(m)}}\right)}, 1 \leq m \leq M, \text{ with} \quad (3)$$

$$\delta^{(k)} = \frac{\beta^{(k)}}{P} \sum_{p=1}^{P} w^{(k,p)} e^{j(\varphi^{(k,p)} + \psi^{(k,p)})}, \quad (4)$$

where $\frac{\beta^{(k)}}{P}$ is the gain (modeled as a complex random gaussian variable under Rayleigh fading assumption, with a total unitary power over the entire RIS, i.e. with $\mathrm{E}\left[\left|\beta^{(k)}\right|^2\right] = 1$) of the path between the 1st antenna element of the BS and the target UE, passing by the 1st element of the $k^{th}$ RIS, $w_{k,p}$ is the phase-shift weight of the $p^{th}$ element of the $k^{th}$ RIS, $\varphi_{k,p}$ and $\psi_{k,p}$ are phase-shifts defined hereafter. $\varphi_{k,p}$ and $\psi_{k,p}$ are given by:

$$\varphi^{(k,p)} = \frac{2\pi}{\lambda}\left(\overrightarrow{a^{(k)}} \cdot \overrightarrow{c^{(k,p)}}\right), \quad (5)$$

$$\psi^{(k,p)} = \frac{2\pi}{\lambda}\left(\overrightarrow{b^{(k)}} \cdot \overrightarrow{c^{(k,p)}}\right), \quad (6)$$

where, $\overrightarrow{c^{(k,p)}}$ is the vector between the positions of the 1st and the $p^{th}$ element of the $k^{th}$ RIS, $\overrightarrow{a^{(k)}}$ and $\overrightarrow{b^{(k)}}$ are the unitary vectors indicating the BS-to-$k^{th}$ RIS direction and the $k^{th}$ RIS-to-target UE direction, respectively.

Finally, we consider a position Q close to the BS. For such position, we assume a free-space propagation and a spherical wave model using Friis' formula. With these assumptions, the channel vector $\mathbf{q}(Q) \in \mathbb{C}^{1 \times M}$ between the BS and the UE Q is given by:

$$\mathbf{q}_m(Q) = \frac{\lambda}{4\pi d^{(m)}(Q)} e^{j2\pi \frac{d^{(m)}(Q)}{\lambda}}, m \leq M. \quad (7)$$

where, $d^{(m)}(Q)$ is the distance between the $m^{th}$ element of the BS and the UE Q.

*B. RIS-assisted BF procedure, received powers at T and Q*

The RIS-assisted BF procedure [11] is illustrated in Fig. 3 and is composed of two phases detailed hereafter. Note that the BF procedure without RIS assistance simply consists of the second phase only.

During the RIS self-configuration phase, the target UE sends pilots in the uplink, to allow any RIS $k$ to estimate the target UE-to-RIS channel phases $\psi^{(k,p)}$'s. We consider a RIS with continuous phase-shifting capability, as the prototypes used in [17][18]. Hence, each RIS can computes weights $w_{k,p}$'s to 'turn itself electronically' towards the target UE, as follows:

$$w^{(k,m,p)} = e^{-j\psi^{(k,p)}}. \quad (8)$$

Then, each RIS reconfigures its weights and 'freezes". Hence, the propagation channel is pre-shaped by RISs for the target UE.

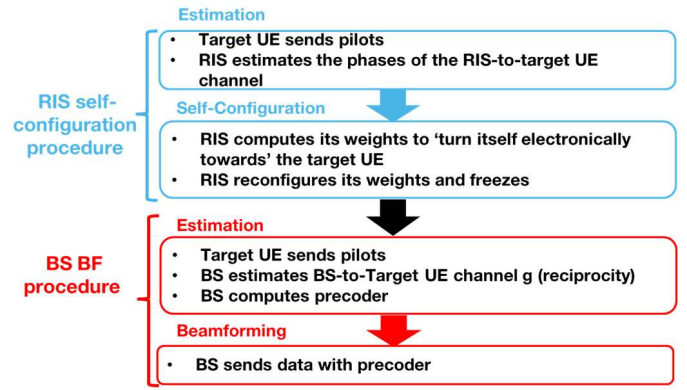

Fig. 4. RIS-assisted Beamforming Procedure

Then, during the BS BF phase, the target UE sends pilots in the uplink again, to allow the BS to measure the target UE-to-BS channel (that has been previously pre-shaped during the RIS self-configuration phase). The BS exploits channel reciprocity and deduces the BS-to-Target UE channel $\mathbf{g}$. Finally, the BS computes the unitary beamforming vector $\mathbf{b} \in \mathbb{C}^{M \times 1}$ (with $\|\mathbf{b}\| = 1$) based on $\mathbf{g}$. The BS transmit power is denoted by $\chi \leq$ $\chi^{max}$. The received power $\rho$ at the target UE T and the received power $\omega(Q, \mathbf{b}, \chi)$ at the UE Q close to the BS are given by:

$$\rho = |\mathbf{gb}|^2 \chi, \quad (9)$$

$$\omega(Q, \mathbf{b}, \chi) = |\mathbf{q}(Q)\mathbf{b}|^2 \chi. \quad (10)$$

Note that the BS does not even need to be aware that there is a RIS.

Depending on the BF scheme, the expressions of $\mathbf{b}$ and $\chi$ differ. We denote by $\mathbf{b}^{MRT}$, $\mathbf{b}^{red}$ and $\mathbf{b}^{eq}$ the precoders for the MRT, the reduced, and the equalized schemes, respectively. We denote by $\chi^{MRT}$, $\chi^{red}$ and $\chi^{eq}$, the transmit powers for the MRT, the reduced and the equalized schemes, respectively.

### C. MRT BF precoder & transmit power

For the MRT BF [3], $\mathbf{b}^{MRT}$ and $\chi^{MRT}$ are simply given by:

$$\mathbf{b}^{MRT} = \frac{\mathbf{g}^\dagger}{\|\mathbf{g}^\dagger\|}, \quad (13)$$

$$\chi^{MRT} = \chi^{max}. \quad (14)$$

### D. Reduced MRT BF precoder & transmit power

For the reduced MRT BF [11]:

$$\mathbf{b}^{red} = \mathbf{b}^{MRT}, \quad (15)$$

$$\chi^{red} = min\left(\frac{\omega^{threshold}}{\omega^{max}} \chi^{max}, \chi^{max}\right). \quad (16)$$

where $\omega^{max}$ is the maximum power received on the limit circle $\mathcal{C}$, and defined by:

$$\omega^{max} = \max_{Q \subset \mathcal{C}} \left\{ \frac{\omega(Q, \mathbf{b}^{red}, \chi^{max})}{\chi^{max}} \right\}. \quad (17)$$

$\omega^{max}$ is determined numerically by the BS, using (10) and (7).

### E. Novel angularly equalized MRT BF precoder & transmit power

The proposed novel Equalized BF scheme is computed by the BS as follows. First, the BS is assumed to have perfect knowledge of the $\overrightarrow{v^{(m)}}$ vectors, indicating the position of the BS antenna element $m$ relatively to the 1st element, for each $m$. Then, using channel sounding methods as those in [19]-[21], the BS estimates the directions of departure $\overrightarrow{u^{(n)}}$, $\overrightarrow{a^{(k)}}$ of the propagation paths between the BS and the scatterer $n$, and between the the BS and the RIS $k$, respectively, for each $n$ and each $k$, respectively. We assume that these estimates are perfect. Then, the BS computes an *equalized virtual propagation channel* $\mathbf{g}' \in \mathbb{C}^{1 \times M}$, as follows:

$$\mathbf{g}' = \mathbf{s}' + \mathbf{h}', \quad (18)$$

where, $\mathbf{s}', \mathbf{h}' \in \mathbb{C}^{1 \times M}$ are defined hereafter:

$$\mathbf{s}'_m = \sum_{n=1}^{N} \alpha'^{(n)} e^{j\frac{2\pi}{\lambda}\left(\overrightarrow{u^{(n)}} \cdot \overrightarrow{v^{(m)}}\right)}, 1 \leq m \leq M, \quad (19)$$

$$\alpha'^{(n)} = 1, 1 \leq n \leq N, \quad (20)$$

$$\mathbf{h}'_m = \sum_{k=1}^{K} \delta'^{(k)} e^{j\frac{2\pi}{\lambda}\left(\overrightarrow{a^{(k)}} \cdot \overrightarrow{v^{(m)}}\right)}, 1 \leq m \leq M, \quad (21)$$

$$\delta'^{(k)} = 1, 1 \leq k \leq K. \quad (22)$$

As illustrated in Fig. 2, compared to the true propagation channel, the angularly equalized virtual propagation channel has all its propagation paths equalized in strength, by setting $\alpha'^{(n)} = \delta'^{(k)} = 1$ for all $n$ and $k$. Finally, the BS computes the equalized BF precoder $\mathbf{b}^{eq} \in \mathbb{C}^{M \times 1}$ as the precoder adapted to the angularly equalized virtual propagation channel $\mathbf{g}'$:

$$\mathbf{b}^{eq} = \frac{\mathbf{g}'^\dagger}{\|\mathbf{g}'^\dagger\|}. \quad (23)$$

The transmit power $\chi^{eq}$ is then deduced by the BS:

$$\chi^{eq} = min\left(\frac{\omega^{threshold}}{\omega'^{max}} \chi^{max}, \chi^{max}\right), \quad (24)$$

where $\omega'^{max}$ is the maximum power received on the limit circle $\mathcal{C}$ and defined by:

$$\omega'^{max} = \max_{Q \subset \mathcal{C}} \left\{ \frac{\omega(Q, \mathbf{b}^{eq}, \chi^{max})}{\chi^{max}} \right\}. \quad (25)$$

$\omega'^{max}$ is determined numerically by the BS using (7) and (10).

## III. VISUALISATION OF THE IMPACT OF BF SCHEMES UPON THE OVER-EXPOSED-AREA

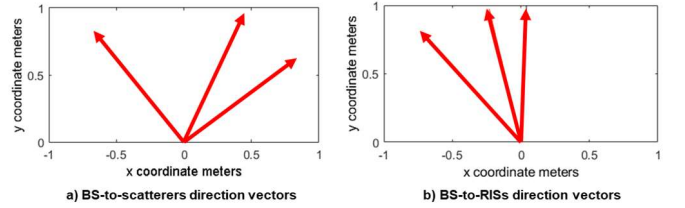

Fig. 5. BS-to-scatterers and BS-to-RIS directions

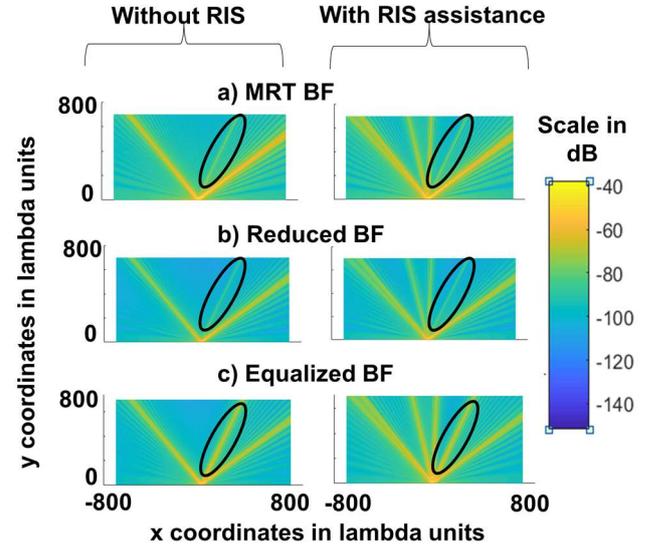

Fig. 6. Temporary received power $\omega$ close to the BS, for all BF schemes (with one weak path being highlighted with a black circle), assuming the same target UE is served for a long period

In this Section, we propose to visualize the impact of the studied BF schemes, over the over-exposed area (i.e., the area where $\omega > \omega^{threshold}$), for a given random propagation channel sample. For each BF scheme with a precoder **b** and a transmit power $\chi$, we calculate the received power $\omega(Q, \mathbf{b}, \chi)$ at locations Q near the BS, using the mathematical expressions given in Section II. The obtained results correspond to one random channel sample. The simulation parameters are fixed as follows: $M = 64$, $N = 3$, $K = 3$, $P = 16$, $R = 650$ (this value being the radius normalized by $\lambda$) and $\frac{\omega^{threshold}}{\chi^{max}} = -70$dB. Throughout this section, all distances and powers are normalized by $\lambda$ and $\chi^{max}$, respectively. We consider two-dimensional (2D) propagation in the x-y plane. The linear array of the BS antenna is at coordinate (0, 0) and is deployed along the x-axis. The BS-to-scatterers and BS-to-RISs directions are plotted in Fig. 5-a) and Fig. 5-b), respectively.

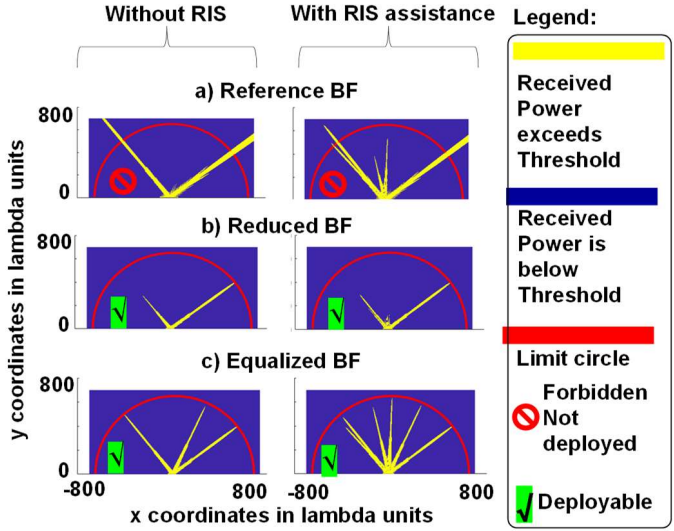

Fig. 7. Over-exposed area (yellow) and limit circle (red) for all BF schemes, assuming the same target UE is served for a long period.

In Fig. 6, to visualize the radiation pattern, we plot the received power $\omega$ (in dB) at positions near the BS for all the studied schemes. In Fig. 7, we plot the over-exposed area (i.e., the area where $\omega > \omega^{threshold}$), in yellow and we plot in red the limit circle $\mathcal{C}$. As expected, as observable on Fig. 6, as they are all derived from MRT BF, all BF schemes radiate towards the scatterers and the RISs, i.e., in the same directions as those depicted in Fig. 5. As we can observe on Fig. 7, the over-exposed areas as well, stretch themselves in the directions of the scatterers and RISs, except for one scatterer. Indeed, the radiation, circled in black in Fig. 6, in the direction of one scatterer, is so weak that it does not appear in the over-exposed areas of the MRT BF and the reduced BF schemes. We can also clearly visualize that:

- the over-exposed area of MRT BF exceeds the limit circle only in certain directions and is inside the circle in others.
- the over-exposed area of Reduced BF is severely reduced in all directions (it is an homothety of the MRT BF over-exposed area) to ensure that the area exactly meets the circle in its strongest direction.
- the Equalized BF scheme reduces the over-exposed area in three directions and stretches the area in three others, to get the area exactly upon the circle in six directions.
- the RISs spread the radiation pattern in the angular domain regardless of the considered BF scheme.

TABLE I. NORMALIZED RECEIVED POWER AT TARGET UE AND PERCENTAGE OF POSITIONS (OUTSIDE THE LIMIT CIRCLE AND INSIDE A $1400\lambda$ BY $1400\lambda$ SQUARE AROUND THE BS) EXCEEDING THE THRESHOLD, ASSUMING THE SAME TARGET UE IS SERVED FOR A LONG PERIOD

| RIS | BF Scheme | Received Power at Target UE (dB) | Percentage of positions in scanned area |
|---|---|---|---|
| No | MRT | 25.2 | 3.8 |
|  | Reduced | 17 | 0 |
|  | Equalized | 18.8 | 0 |
| Yes | MRT | 26.5 | 4.5 |
|  | Reduced | 19.5 | 0 |
|  | Equalized | 23.1 | 0 |

Table I reports the received power at the target UE and the percentage of positions exceeding the $\omega^{threshol}$ beyond the limit circle in square area of $1400\lambda$ by $1400\lambda$ around the BS. Table I shows that Equalized BF maximizes the received power at the target UE, whilst complying with the EMFE constrain. Table I also shows the benefit of RIS assistance.

IV. STATISTICAL RESULTS

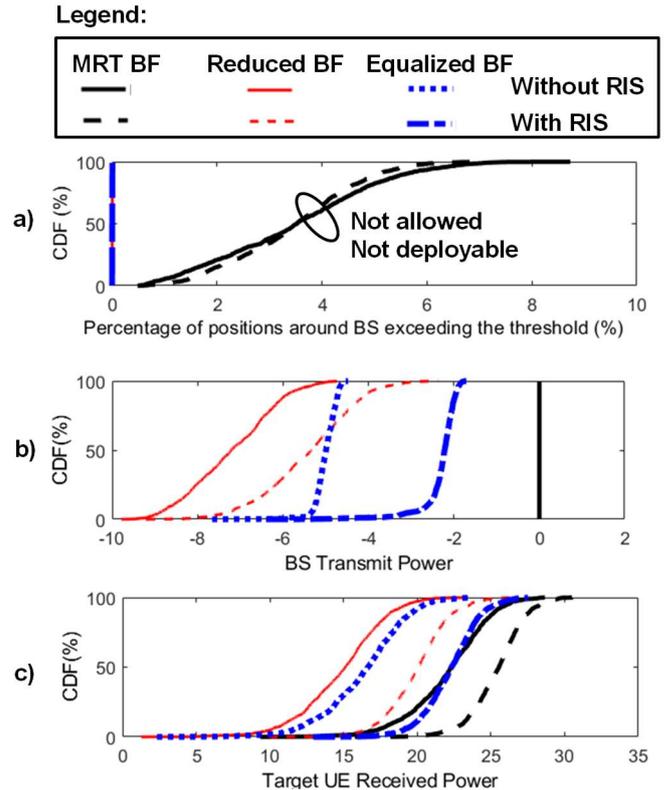

Fig. 8. CDFs of a) positions around BS (outside the limit circle and inside a $1400\ \lambda$ by $1400\ \lambda$ square around the BS) exceeding the threshold b) BS transmit power and c) target UE received power, assuming the same target UE is served for a long period.

In this section, the performance of each BF scheme is evaluated statistically over random channels. This time, we draw $Nsamples = 1000$ random channel samples. For each sample and for each scheme, we compute (using the model described in Section II) the following three metrics: the percentage of positions with the scanned area (i.e., beyond the safety circle and inside a $1400\lambda \times 1400\lambda$ square around the BS) with a received power exceeding the threshold; the transmit power $\chi$ and the received power $\omega$ at the target UE. Then, the cumulative density function (CDF) is computed over all samples, for each metric. The simulation parameters are fixed as follows: $M = 64$, $N = 3$, $K = 3$, $P = 16$, $R = 650$ (where $R$ is normalized by $\lambda$) and $\frac{\omega^{threshold}}{\chi^{max}} = -70$dB. Again, all distances and powers are normalized by $\lambda$ and $\chi^{max}$, respectively. The BS linear antenna array is along the x-axis. The directions from the BS to the scatterers and RISs are randomly distributed. Fig. 8-a), Fig. 8-b) and Fig. 8-c) present the CDFs of the previously mentioned metrics. The most significant received power on the target UE is obtained with the MRT BF scheme with a violation of the EMFE constraint. On the contrary, this constrain is perfectly met by the Reduced BF scheme, but at the expense of the received power at the target UE, due to a weaker BS transmit power. The Equalized BF scheme improves the received power (compared to the Reduced BF) whilst meeting the EMFE constrain. One can also notice that this received power gain comes at the expense of a large transmit power increase. Indeed, the BS must strongly boost its power in the direction of weak propagation paths.

As a conclusion, Equalized BF scheme outperforms other schemes, in terms of received power at the target UE, under EMFE constrain.

## V. CONCLUSION

In this paper, we propose a novel beamforming scheme called "Equalized" beamforming assisted by self-tuned reconfigurable intelligent surfaces. Compared to the standard maximum ratio transmission beamforming, the proposed new system modifies its radiation pattern in such way that the power delivered to the target user equipment is increased, whilst the over-exposed area (the area where the received power exceeds the threshold fixed by the regulation) remains inside a pre-defined limit circle. More precisely, the (radiation pattern) and the over-exposed area is (i) spread in the angular domain, to provide additional paths towards the target user equipment, (ii) reduced in directions exceeding the circle and (iii) stretched in directions inside the circle. Our simulations show that Equalized beamforming maximizes the received power under exposure constrain. Next studies will extend the use of the angularly equalized virtual channel to other beamformers than maximum ratio transmission.

## ACKNOWLEDGMENT

This work was partially conducted within the framework of the European Union's Horizon 2020 research and innovation project RISE-6G under EC Grant 101017011. We thank our colleagues at Orange, Yuan-Yuan Huang and Dominique Nussbaum for their feedbacks.